\documentclass{nature_meth}
\usepackage{amssymb,amsfonts,amsmath}
\usepackage{graphicx} 
\usepackage{gensymb}
\usepackage{rotating}

 \usepackage{multirow}
\usepackage{epsfig}
\usepackage[figoff]{figcaps}
\usepackage{hyperref}
\hypersetup{
    colorlinks=true,
    linkcolor=blue,
    filecolor=magenta,
    urlcolor=cyan,
}

\usepackage{bm} 
\usepackage{xspace}

\newcommand{\xmath}[1] {\ensuremath{#1}\xspace}
\newcommand{\blmath}[1] {\xmath{\bm{#1}}}

\newcommand{\Ab}{\blmath{A}}
\newcommand{\Bb}{\blmath{B}}
\newcommand{\Cb}{\blmath{C}}

\newcommand{\Ib}{\blmath{I}}

\usepackage{caption,setspace}
\renewcommand{\spacing}[1]{\renewcommand{\baselinestretch}{#1}\large\normalsize}
\spacing{2}

\usepackage{url}

\usepackage{graphicx}
\makeatletter
\let\saved@includegraphics\includegraphics
\AtBeginDocument{\let\includegraphics\saved@includegraphics}
\renewcommand{\figurename}{Fig.}
\renewcommand{\thefigure}{\arabic{figure}}
\makeatother

\title{scMamba: A Pre-Trained Model for Single-Nucleus RNA Sequencing Analysis in Neurodegenerative Disorders}

\author{Gyutaek Oh$^{1, \ast}$,
        Baekgyu Choi$^{2, \ast}$,
        Seyoung Jin$^{2}$,
        Inkyung Jung$^{2, \dagger}$,
        and Jong Chul Ye$^{3, \dagger}$}

\begin{document}
\setstretch{1.2}

\maketitle

\begin{affiliations}
\item Department of Bio and Brain Engineering, Korea Advanced Institute of Science and Technology (KAIST), Daejeon, Republic of Korea
\item Department of Biological Sciences, Korea Advanced Institute of Science and Technology (KAIST), Daejeon, Republic of Korea
\item Kim Jaechul Graduate School of AI, Korea Advanced Institute of Science and Technology (KAIST), Daejeon, Republic of Korea
\item[] $^{\dagger}$Correspondence should be addressed to Jong Chul Ye (jong.ye@kaist.ac.kr) or Inkyung Jung (ijung@kaist.ac.kr).
\end{affiliations}
 
\vspace{-0.5cm}
\section*{Abstract}
\begin{abstract}
Single-nucleus RNA sequencing (snRNA-seq) has significantly advanced our understanding of the disease etiology of neurodegenerative disorders.
However, the low quality of specimens derived from postmortem brain tissues, combined with the high variability caused by disease heterogeneity, makes it challenging to integrate snRNA-seq data from multiple sources for precise analyses.
To address these challenges, we present scMamba, a pre-trained model designed to improve the quality and utility of snRNA-seq analysis, with a particular focus on neurodegenerative diseases.
Inspired by the recent Mamba model, scMamba introduces a novel architecture that incorporates a linear adapter layer, gene embeddings, and bidirectional Mamba blocks, enabling efficient processing of snRNA-seq data while preserving information from the raw input.
Notably, scMamba learns generalizable features of cells and genes through pre-training on snRNA-seq data, without relying on dimension reduction or selection of highly variable genes.
We demonstrate that scMamba outperforms benchmark methods in various downstream tasks, including cell type annotation, doublet detection, imputation, and the identification of differentially expressed genes.

\end{abstract}

\clearpage
\setstretch{1.6}

\section*{Introduction}
Single-cell RNA sequencing (scRNA-seq) is a powerful technique for profiling gene expression at single-cell resolution, enabling the exploration of molecular characteristics in complex biological systems under both normal and disease conditions \cite{saliba2014single,rood2022impact}.
scRNA-seq analysis facilitates several key objectives, including cell type annotation \cite{li2020scibet, hao2021integrated}, the discovery of novel cell types \cite{villani2017single}, the identification of marker genes \cite{jaitin2014massively}, and the analysis of cellular heterogeneity \cite{papalexi2018single,kinker2020pan}.

Notably, the brain exhibits an exceptionally diverse range of cell types compared to other tissues \cite{saunders2018molecular,hodge2019conserved}.
As a result, single-cell RNA sequencing in the brain is crucial for gaining deeper insights into brain function within various cellular contexts.
Due to the highly interconnected nature of brain tissue, isolating nuclei for RNA sequencing—known as single-nucleus RNA sequencing (snRNA-seq)—is a more commonly used approach in brain research than scRNA-seq.
Recent studies utilizing snRNA-seq in neurodegenerative diseases, such as Parkinson’s disease and Alzheimer’s disease, have uncovered disease-vulnerable subtypes of neurons and associated glial cell subtypes, shedding light on the heterogeneity and complexity of the cellular landscape in neurodegenerative disorders \cite{keren2017unique,mathys2019single,habib2020disease,leng2021molecular,smajic2022single}.

However, snRNA-seq analysis in neurodegenerative diseases faces several significant challenges.
First, the low quality of postmortem brain samples, which are obtained from brains at varying postmortem intervals, often results in poor-quality snRNA-seq data.
Second, the high variability of snRNA-seq, compounded by disease heterogeneity, makes it even more difficult to integrate data from multiple sources.
Finally, the limited quantity of mRNA within a single nucleus increases the likelihood of gene expression failure, a phenomenon known as ``dropout''.
As a result, snRNA-seq data often contains a high number of zero counts, making it essential to differentiate between true biological zeros and false zeros caused by technical noise.

To address these challenges, several computational approaches have been developed, primarily targeting single-cell RNA sequencing.
However, two critical issues remain in previously developed methods.
First, numerous imputation techniques have been introduced to resolve dropout events in scRNA-seq data \cite{huang2018saver,li2018accurate,van2018recovering,arisdakessian2019deepimpute,eraslan2019single}.
However, many of these existing methods suffer from long computational runtimes, underscoring the need for a more efficient approach to impute missing values in scRNA-seq data.

Another challenge stems from the long sequence length of snRNA-seq data.
Typically, snRNA-seq captures the expression levels of tens of thousands of genes, and cell type annotation methods \cite{aran2019reference,li2020scibet,hao2021integrated,yang2022scbert} classify cell types based on gene expression patterns.
However, analyzing information from all genes is computationally demanding and often exceeds the capacity of many models.
To address this, many annotation methods focus on highly variable genes (HVGs), a subset of a few thousand genes, and rely exclusively on their expression levels.
Unfortunately, the selection of HVGs is sensitive to parameter settings and can vary significantly across datasets and batches (e.g., between different patients).
Moreover, selecting too few HVGs risks losing critical cellular information.
This underscores the need for analysis methods capable of utilizing information from all genes without relying on HVG selection.

In recent years, pre-trained models have gained significant attention and been applied across various data types \cite{devlin2018bert,brown2020language,bommasani2021opportunities,ramesh2021zero}.
These models typically undergo pre-training using large, unlabeled datasets via self-supervised learning to extract generalizable features within a specific data domain.
After pre-training, these foundation models can be fine-tuned with smaller labeled datasets, enabling their effective application to a wide range of downstream tasks.

Inspired by the success of pre-trained models, several pre-trained models for scRNA-seq have been proposed \cite{yang2022scbert,theodoris2023transfer,cui2024scgpt,hao2024large}.
Some of these methods utilize Transformer architectures, where computational complexity increases quadratically with the input length.
To address this, they often rely on a subset of genes to reduce computational demands.
Other methods discretize expression values into bins, treating scRNA-seq data similarly to tokens in language models.
While these strategies effectively manage processing requirements, they risk introducing significant information loss in the scRNA-seq data.

\begin{figure}[!t]
    \centering
    \includegraphics[width=0.99\linewidth]{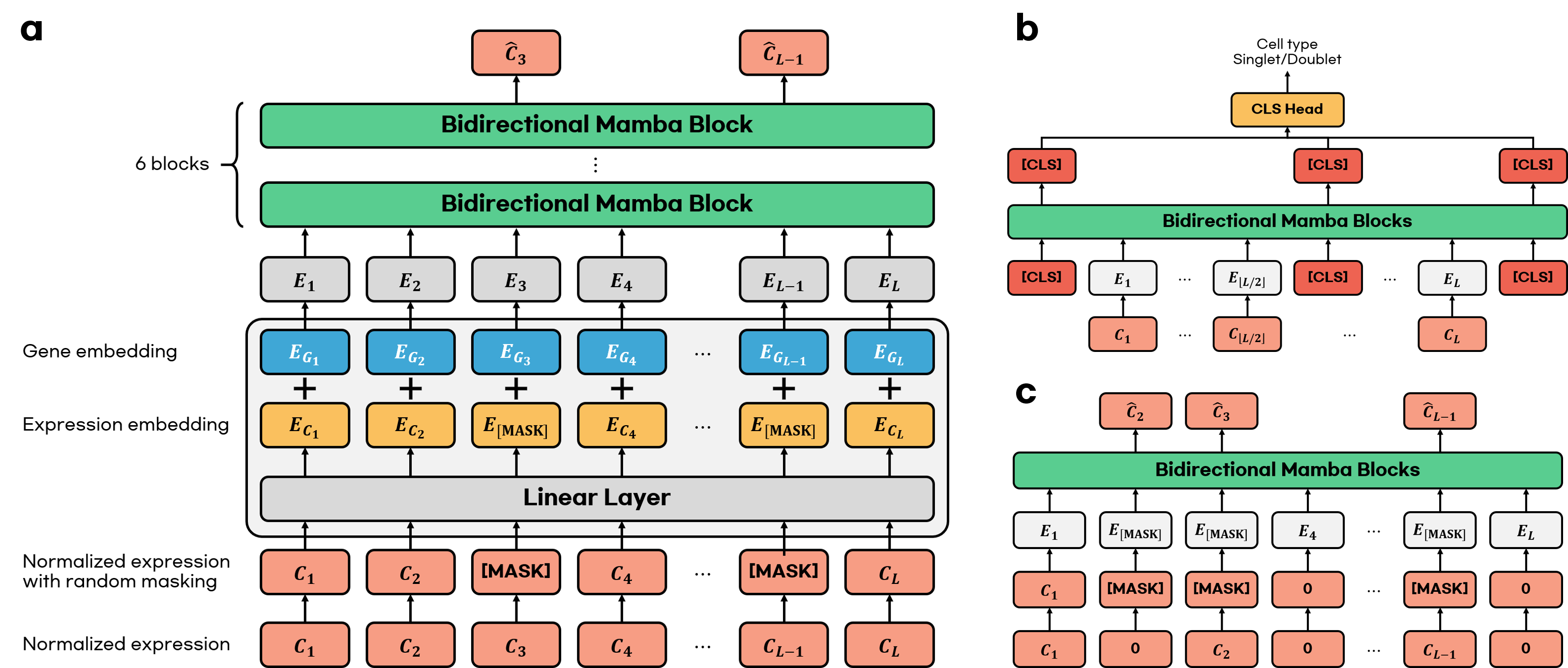}
    \caption{\bf\footnotesize
    Overall framework of scMamba.
    \textbf{a}. The scMamba model comprises a linear layer for expression embeddings, gene embeddings, and bidirectional Mamba blocks.
    During pre-training, a subset of input data is masked, and the model predicts the expression levels at the masked positions.
    \textbf{b}. For fine-tuning classification tasks, three [CLS] embeddings are inserted into the input embeddings.
    These embeddings are processed through the Mamba blocks and then passed to a classification head, which predicts cell classes.
    \textbf{c}. For fine-tuning the snRNA-seq imputation task, portions of the input expression levels are masked, and the model predicts the masked values.
    Zero and non-zero values are masked with different probabilities to ensure balanced learning.
    Both zero and non-zero values are masked with different masking probabilities.
    }
    \label{fig:scmamba}
\end{figure}

Recently, a novel architecture called Mamba \cite{gu2024mamba} was introduced.
Mamba is based on selective state space models (SSMs), enabling it to select data in an input-dependent manner dynamically.
It also offers lower computational complexity than Transformers with self-attention, allowing for faster processing of long sequences.
Mamba has demonstrated superior performance over Transformers and other SSM-based architectures in specific tasks and has been successfully applied across various domains, achieving strong results \cite{huang2024mambamir, liu2024vmamba, qiao2024vl, schiff2024caduceus, zhu2024vision, guo2025mambair}.

Inspired by this, we propose scMamba (Fig. \ref{fig:scmamba}a), an enhanced pre-trained model for analyzing brain snRNA-seq data.
scMamba incorporates the Mamba block instead of self-attention, enabling it to process long snRNA-seq data without requiring dimensionality reduction.
By pre-training the model using masked expression modeling, we demonstrate its effectiveness in downstream tasks such as cell type classification, doublet detection, and snRNA-seq imputation (Fig. \ref{fig:scmamba}b, c).
Moreover, we show that scMamba consistently outperforms comparative methods across five diverse datasets from different brain tissues.

\section*{Results}
\subsection{scMamba learn meaningful representation during pre-training.}
\begin{figure}[!t]
    \centering
    \includegraphics[width=0.99\linewidth]{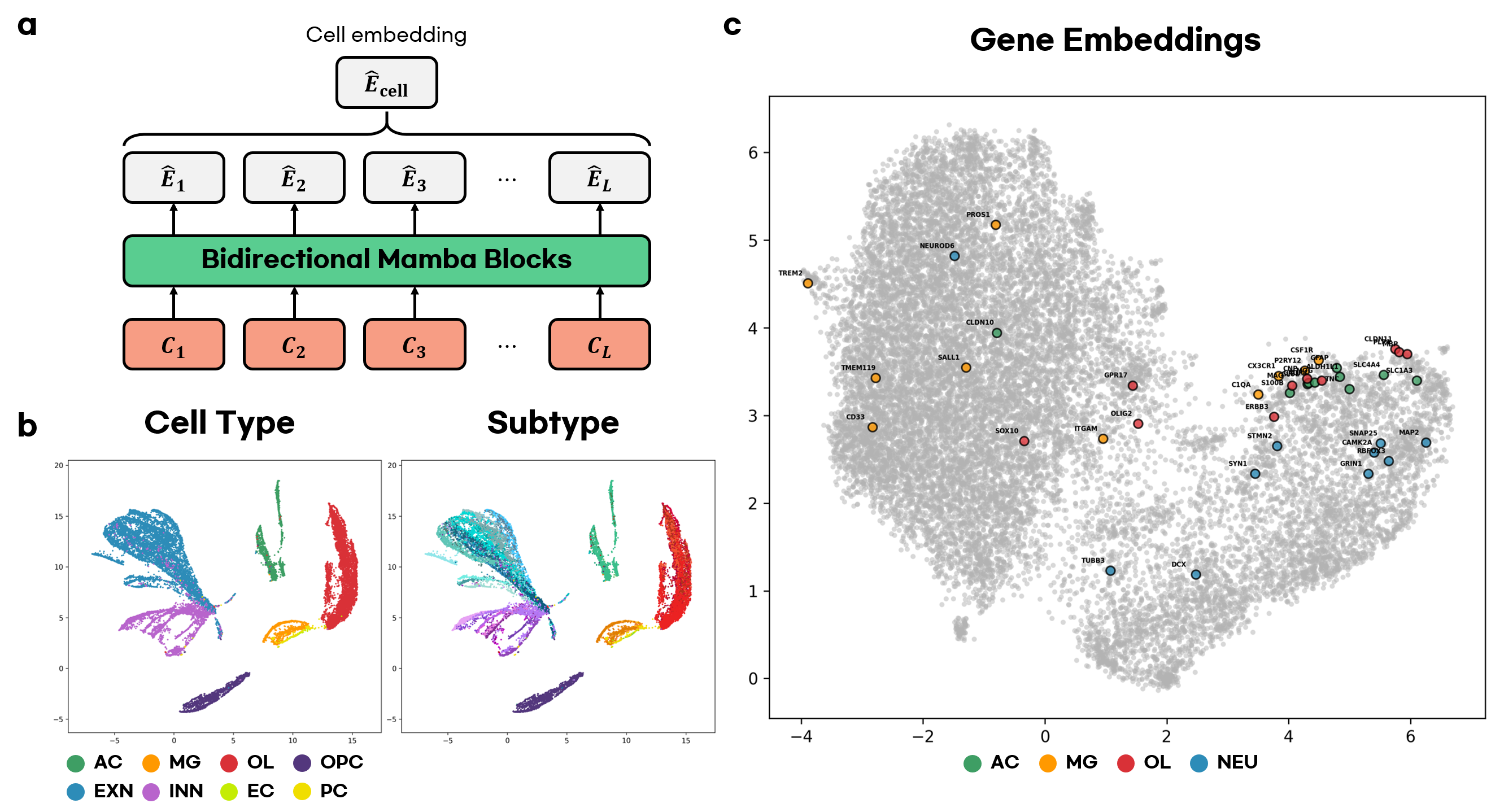}
    \caption{\bf\footnotesize
    \textbf{a}. To generate cell embeddings from the pre-trained model, snRNA-seq data is input into the model, and the resulting output features are averaged along the sequence length.
    \textbf{b}. UMAP visualization of cell embeddings from the pre-trained scMamba model.
    Each UMAP is colored based on 8 major cell types or 72 subtypes. 
    \textbf{c}. UMAP visualization of gene embeddings from pre-trained scMamba model.
    Marker genes of 4 distinct cell types are labeled with names.
    (AC: astrocyte, MG: microglia, OL: oligodendrocyte, OPC: oligodendrocyte progenitor cell, EXN: excitatory neuron, INN: inhibitory neuron, EC: endothelial cell, PC: pericyte, NEU: neuron).
    }
    \label{fig:embeddings}
\end{figure}

To validate that scMamba learns meaningful cell representations during pre-training, we extracted cell features from the pre-trained model and visualized them in 2D space using UMAP \cite{mcinnes2018umap}.
Specifically, we obtained the output of the pre-trained model with dimensions $L \times D$, where $L$ represents the length of snRNA-seq data and $D$ denotes the embedding dimension of the model.
We then calculated the average along the length dimension, which yielded cell embeddings of dimension $D$ for each cell (Fig. \ref{fig:embeddings}a).

Fig. \ref{fig:embeddings}b presents UMAP visualizations of cell embeddings from the \textit{Lau} dataset.
Notably, the cell embeddings generated by scMamba form distinct clusters corresponding to cell types, despite the absence of cell type labels during pre-training.
Furthermore, cells belonging to the same subtypes are predominantly grouped within these clusters.
The results of other datasets can be found in Supplementary Fig. \ref{fig:embeddings_supp}.
This highlights the ability of scMamba to capture meaningful cell representations during pre-training.

Next, we validate that the gene embeddings learned by the pre-trained scMamba model capture meaningful information.
Fig. \ref{fig:embeddings}c shows a UMAP visualization of the gene embeddings learned by scMamba, as illustrated in Fig. \ref{fig:scmamba}a.
In this visualization, the embeddings of marker genes associated with specific cell types are positioned near one another.
This indicates that scMamba successfully learns relationships among genes through its gene embeddings during pre-training.

\subsection{scMamba is capable of classifying sub cell types.}
\begin{figure}[!t]
    \centering
    \includegraphics[width=0.99\linewidth]{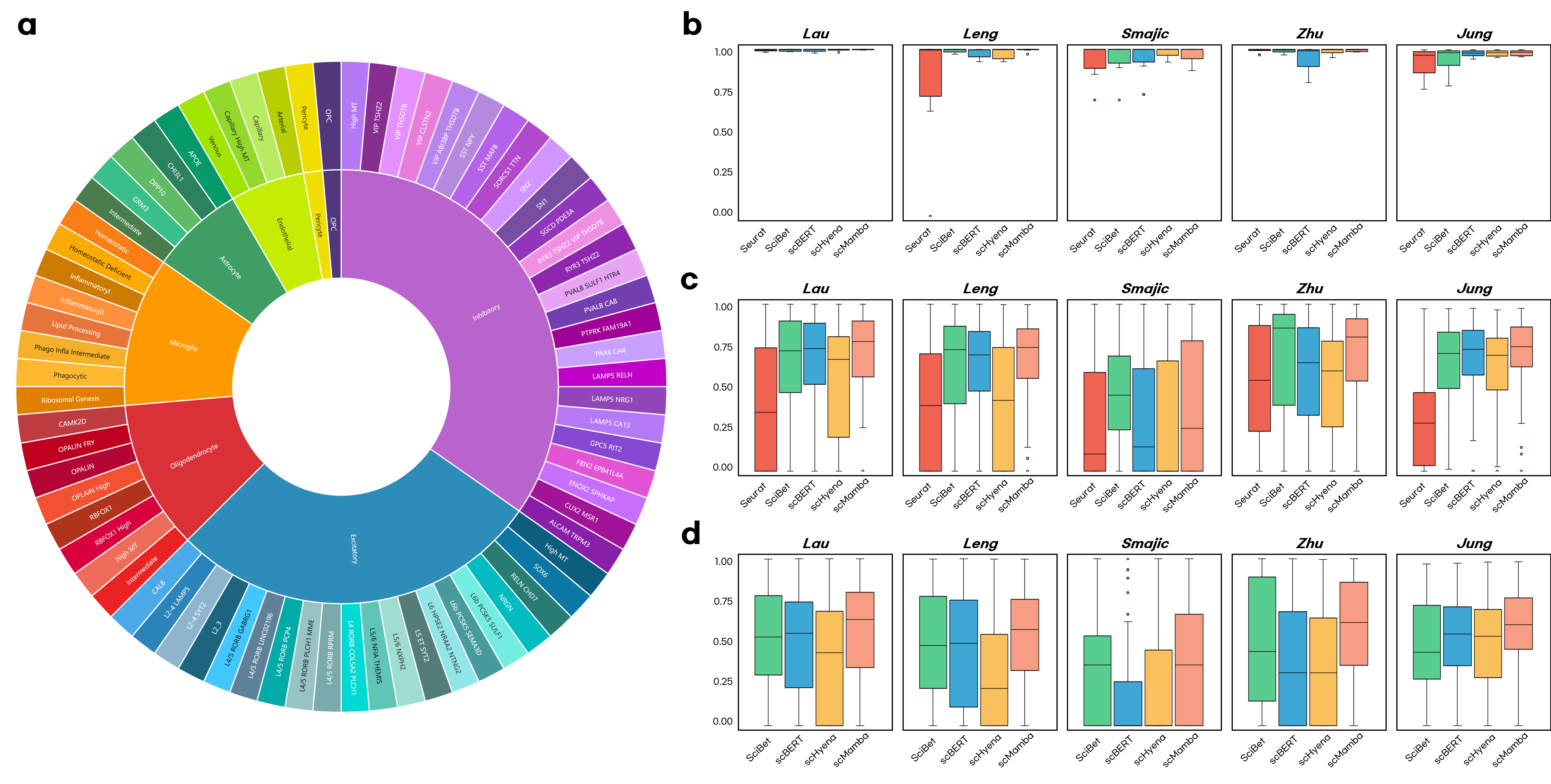}
    \caption{\bf\footnotesize
    \textbf{a}. The dataset is labeled with 8 major cell types and 72 detailed subtypes.
    \textbf{b}. F1 score distribution across 8 cell types, with each box plot representing results for individual datasets.
    \textbf{c}. F1 score distribution across 72 subtypes, with each box plot representing results for individual datasets.
    \textbf{d}. F1 score distribution across 127 subclusters, with each box plot representing results for individual datasets.
    }
    \label{fig:celltype}
\end{figure}

For the cell type classification tasks, our dataset was categorized into 8 major cell types.
Additionally, the dataset was further subdivided into 72 subtypes and 127 fine-grained subclusters.
Fig. \ref{fig:celltype}a provides an overview of the cell types and subtypes included in our dataset.
We conducted cell type classification across three hierarchical levels.

To evaluate the performance of scMamba in the cell type classification task, we performed a comparative analysis against four baseline methods: Seurat \cite{hao2021integrated}, SciBet \cite{li2020scibet}, scBERT \cite{yang2022scbert}, and scHyena \cite{oh2023schyena}.
First, Seurat, a widely used tool for single-cell analysis, was employed.
Using Seurat, we clustered the snRNA-seq data and manually assigned cell types to each cluster by comparing the marker genes of the clusters with known marker genes for each cell type.
Next, SciBet, a supervised cell type annotation method specifically designed for scRNA-seq data, was trained using the same training set utilized for fine-tuning scMamba.
scBERT, another baseline, is a pre-trained model designed for cell type annotation of scRNA-seq data, based on the Performer architecture \cite{choromanski2021rethinking}.
A key difference between scBERT and scMamba is that scBERT discretizes expression levels into bins, while scMamba processes expression levels in their continuous form.
Finally, we included scHyena as a baseline method.
scHyena is another pre-trained model for snRNA-seq data analysis using Hyena \cite{poli2023hyena} operator instead of self-attention.
For a fair comparison, we modified scHyena to use three [CLS] embeddings, as in scMamba, even though the original scHyena implementation uses only a single prepended [CLS] embedding.
To ensure the reproducibility of results on our datasets, we referred to the official source codes of these baseline methods.
Additionally, scBERT and scHyena were pre-trained on the same dataset used for pre-training scMamba.

Fig. \ref{fig:celltype}b illustrates the F1 score distribution across 8 major cell types.
Both scHyena and scMamba achieve consistently high F1 scores across all cell types in each dataset.
Moreover, as shown in Supplementary Table \ref{tab:celltype_supp}, scHyena and scMamba generally achieve the highest average F1 scores across most experiments.
However, other baseline methods also demonstrate strong performance, indicating that classifying major cell types is a relatively straightforward task for classification methods.

To evaluate the classification methods more comprehensively, we applied them to subtype and subcluster classification tasks.
Fig. \ref{fig:celltype}c depicts the F1 score distributions for each classification method.
As shown, Seurat consistently exhibits the lowest performance across all datasets.
Since Seurat assigns clusters to cell types by manually comparing the marker genes of clusters and cell types, its accuracy is highly dependent on the selection of marker genes.
Furthermore, the manual nature of Seurat's classification process demands significant time and effort, which can also affect its overall accuracy.

SciBet performs relatively strongly in subtype classification but falls short compared to the proposed method.
scBERT performs well in most cases, however, its performance declines with the \textit{Smajic} and \textit{Zhu} datasets.
While scHyena excels in major cell type classification, its performance is more constrained in subtype classification, highlighting certain limitations in identifying detailed cell types.

In contrast, the proposed scMamba consistently delivers high performance across all datasets, as confirmed by Supplementary Table \ref{tab:celltype_supp}.
Moreover, both Fig. \ref{fig:celltype}d and Supplementary Table \ref{tab:celltype_supp} demonstrate that scMamba achieves the highest F1 scores in subcluster classification.
These results highlight the superior ability of scMamba to classify detailed cell types more accurately than other methods.

\subsection{scMamba effectively identifies and filters doublets.}
\begin{figure}[!t]
    \centering
    \includegraphics[width=0.99\linewidth]{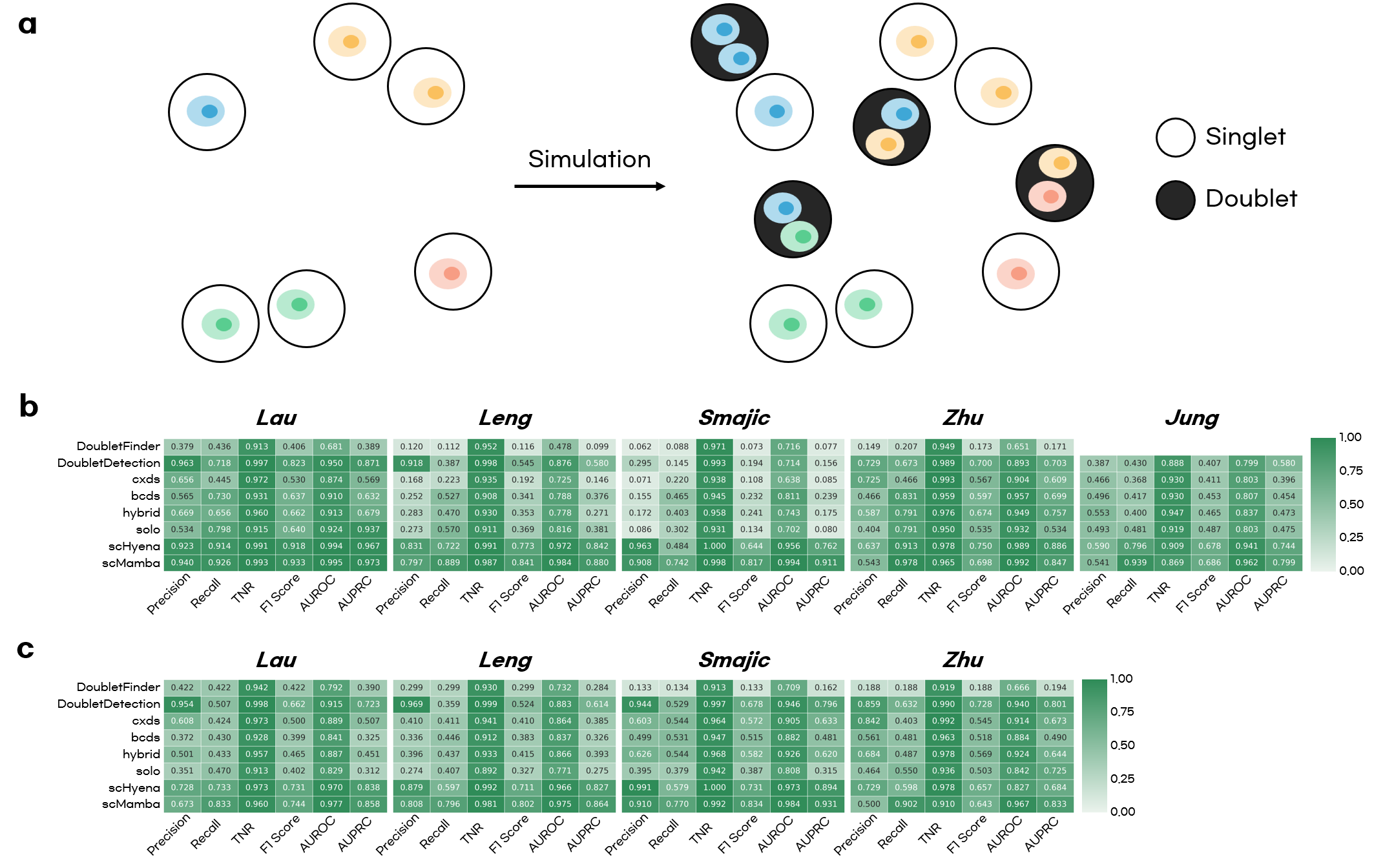}
    \caption{\bf\footnotesize
    \textbf{a}. Simulated doublets are generated by averaging the UMI counts of two randomly selected singlets.
    \textbf{b}. Heatmap of evaluation metric scores for in vivo doublet detection by each method across datasets.
    White squares indicate where the method failed to execute.
    \textbf{c}. Heatmap of evaluation metric scores for simulated doublet detection by each method across datasets.
    }
    \label{fig:doublet}
\end{figure}

During data preparation, we annotated doublets using Scrublet \cite{wolock2019scrublet} to establish ground truth for doublet detection.
However, the labels generated by Scrublet may not always be accurate.
To ensure a fairer comparison, we also conducted experiments using simulated doublets.
As illustrated in Fig. \ref{fig:doublet}a, simulated doublets were created by randomly selecting two singlets and averaging their UMI counts.
The number of simulated doublets was set to 10\% of the total number of singlets.
Experiments using Scrublet-annotated doublets are referred to as `in vivo' doublet detection, while those using simulated doublets are referred to as `simulated' doublet detection.

We followed the approach outlined in the previous study \cite{xi2021benchmarking} to evaluate doublet detection performance.
For baseline comparisons, we utilized seven methods, including scHyena: DoubletFinder \cite{mcginnis2019doubletfinder}, DoubletDetection \cite{gayoso2020doubletdetection},  cxds, bcds, and hybrid from the scds \cite{bais2020scds}, as well as solo \cite{bernstein2020solo}.
All baseline methods were implemented using their official source code.
To assess doublet detection performance, we employed six widely used metrics: precision, recall, true negative rate (TNR), F1 score, area under the ROC curve (AUROC), and area under the precision-recall curve (AUPRC).
Additionally, following the approach described in the previous study \cite{xi2021benchmarking}, we evaluated doublet detection performance under a predefined percentage of droplets identified as doublets (identification rate).

Fig. \ref{fig:doublet}b, c present the evaluation metric scores of each method across datasets for in vivo and simulated doublet detection, respectively.
DoubletFinder consistently demonstrates the lowest performance, particularly with low precision, recall, and F1 scores across all experiments.
Its high computational demands render it unsuitable for larger datasets, such as the \textit{Jung} dataset.
DoubletDetection generally achieves high precision and TNR, effective identifying doublets while minimizing false positives (i.e., singlets misclassified as doublets).
However, it does not surpass the proposed method in other metrics.
While cxds performs poorly in in vivo experiments but improves with simulated data, bcds shows the opposite trend.
Despite these variations, both methods exhibit low overall metric scores.
Similarly, the hybrid method, which combines cxds and bcds, delivers performance comparable to its component methods, while solo achieves slightly lower results.
In contrast, scHyena and scMamba consistently deliver strong performance in doublet detection.
Notably, scMamba achieves the highest recall, F1 score, AUROC, and AUPRC in most experiments.
These results confirm that scMamba is highly effective for accurate doublet detection.

Supplementary Fig. \ref{fig:doublet_invivo_supp} and \ref{fig:doublet_simul_supp} present evaluation metric scores for in vivo and simulated doublet detection, respectively.
Across most experiments with specific identification rates, scMamba consistently achieves the highest precision, recall, and TNR.
These findings further demonstrate the robust doublet detection capability of the proposed method.

\subsection{scMamba can impute snRNA-seq data and correct batch effects.}
\begin{figure}[!t]
    \centering
    \includegraphics[width=0.9\linewidth]{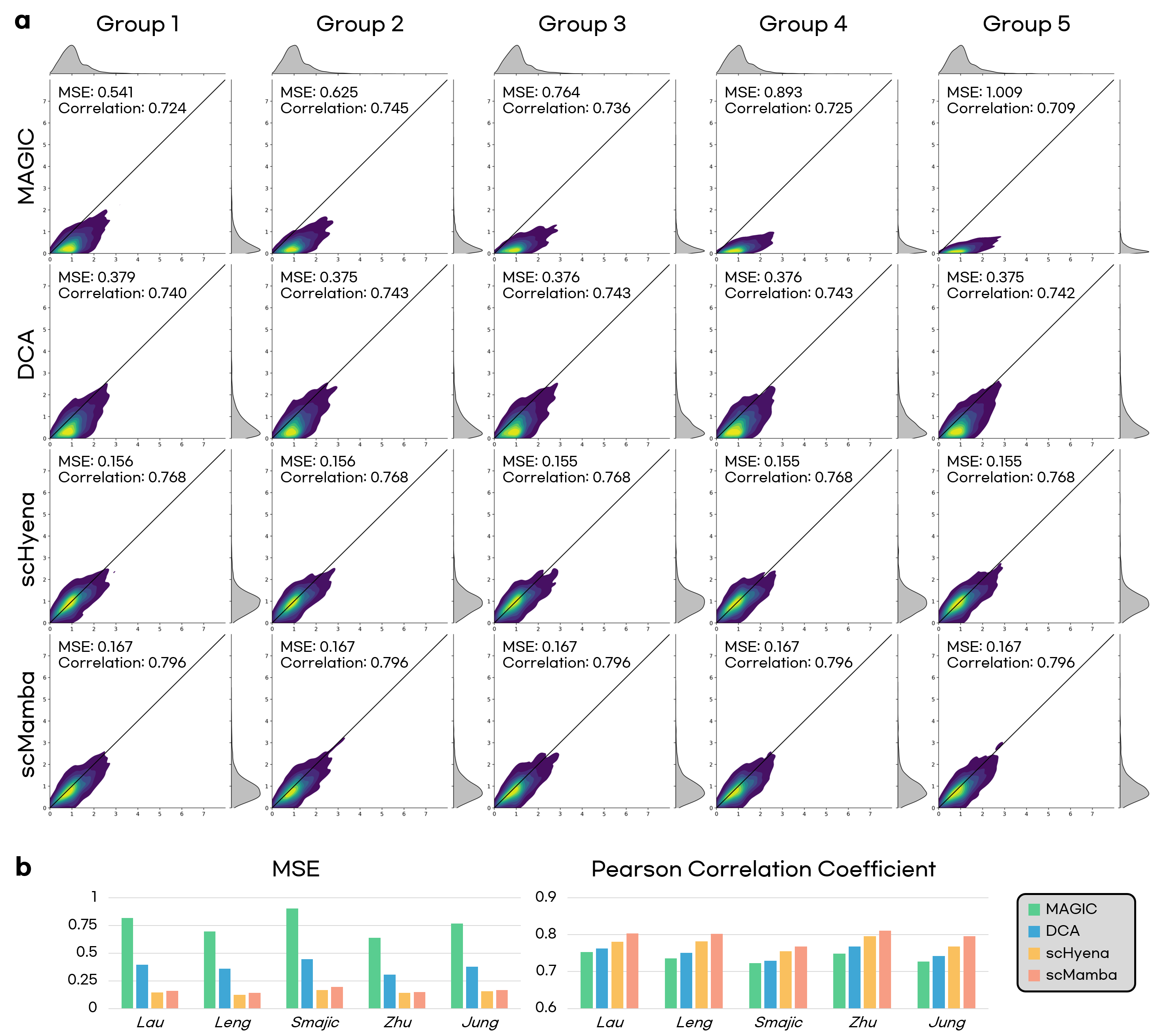}
    \caption{\bf\footnotesize
    Results of nonzero imputation experiments.
    \textbf{a}. Joint plots comparing true and imputed values predicted by various imputation methods on the \textit{Jung} dataset.
    The x and y-axis represent true values and imputed values, respectively.
    Values in the upper left corner of each joint plot indicate the MSEs and Pearson correlation coefficients.
    \textbf{b}. MSE and Pearson correlation coefficient for nonzero imputation by each method across datasets.
    The histogram values represent the averages of the five groups.
    }
    \label{fig:imputation_nonzero}
\end{figure}

To evaluate the imputation performance of scMamba, we compared it with baseline methods MAGIC \cite{van2018recovering}, DCA \cite{eraslan2019single}, and scHyena.
For a quantitative assessment, we first employed a `nonzero imputation' approach: nonzero values in the input data were masked, and the imputation methods were used to predict these masked values.
To ensure comprehensive evaluation, we divided the nonzero indices into five subgroups and evaluated the imputation methods independently on each subgroup.
Performance was measured using the Mean Squared Error (MSE) and Pearson correlation coefficient between the true and imputed values at the masked indices.
Only data with a UMI count greater than 4,000 were included for accurate evaluation.

Fig. \ref{fig:imputation_nonzero}a displays joint plots comparing true and imputed values, along with MSEs and Pearson correlation coefficients for each subgroup in the \textit{Jung} dataset.
The results show that scHyena and scMamba outperform MAGIC and DCA methods in both MSE and Pearson correlation coefficients.
Specifically, MAGIC and DCA tend to underestimate values relative to the true values.
In contrast, the joint plots for scMamba reveal that most points closely align with the $y=x$ line, indicating a strong agreement between true and imputed values.
These results demonstrate that scMamba provides more accurate imputations with significantly lower errors than MAGIC and DCA.
Fig. \ref{fig:imputation_nonzero}b presents histograms of the average MSE and Pearson correlation coefficient for nonzero imputation by each method across datasets.
Once again, scHyena and scMamba exhibit low MSEs and high Pearson correlation coefficients, highlighting their superior performance.
In contrast, MAGIC and DCA demonstrate comparatively poorer results, indicating their limitations in imputation tasks.

\begin{figure}[!t]
    \centering
    \includegraphics[width=0.99\linewidth]{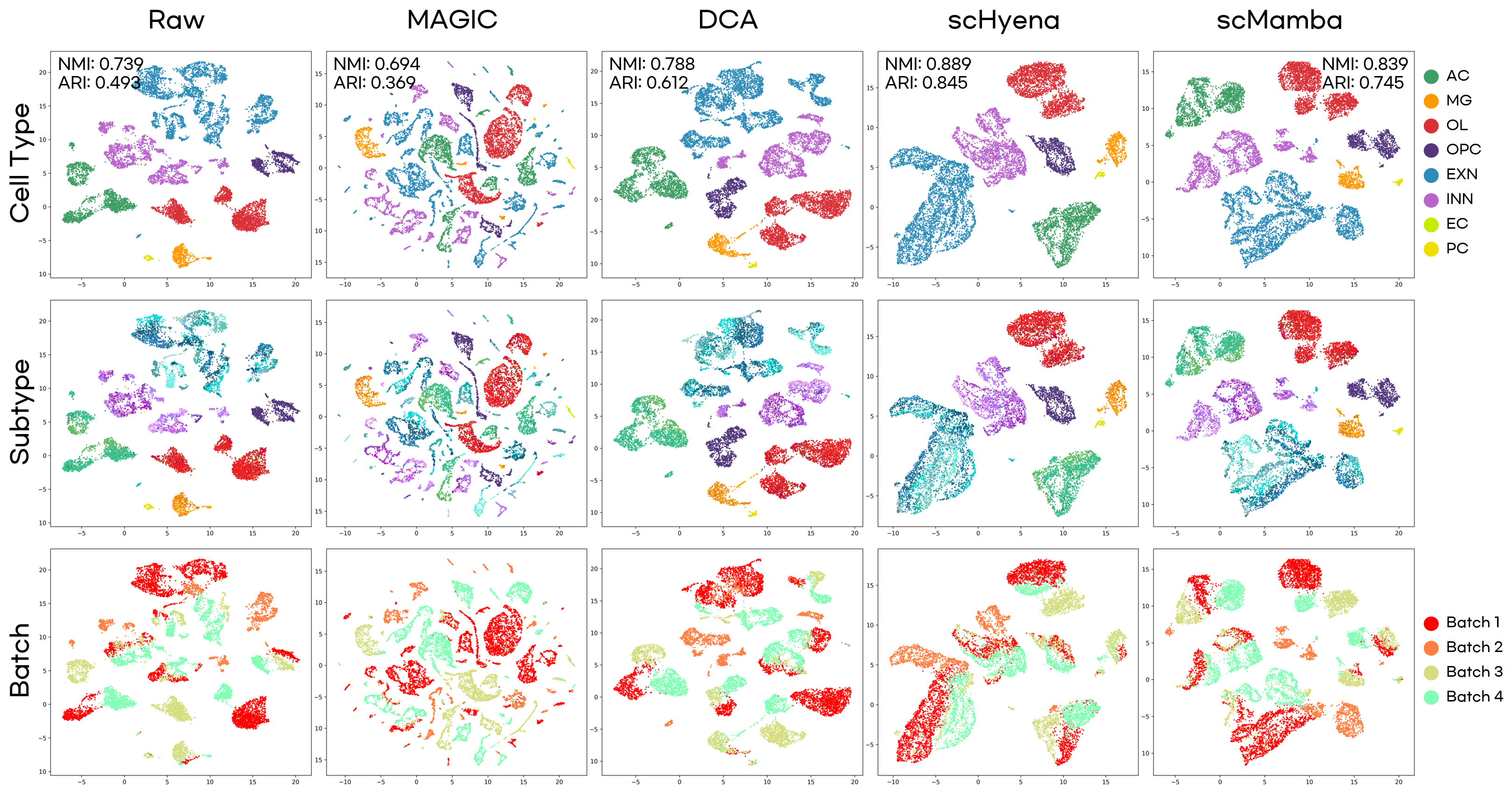}
    \caption{\bf\footnotesize
    Results of zero imputation experiments.
    UMAP visualizations of raw and imputed snRNA-seq data for the \textit{Leng} dataset.
    The figures are labeled with cell type, subtype, and batch (patient), respectively.
    }
    \label{fig:imputation_zero}
\end{figure}

To further evaluate the imputation performance of scMamba, we conducted a `zero imputation` analysis, where zero values in the snRNA-seq data were imputed using various methods and results were visualized in 2D space using UMAP.
Theoretically, accurate imputation of zero values should lead to denser clusters, grouping samples more distinctly with others of the same cell type.
As a result, imputation can help reduce batch effects in the data.
To quantitatively evaluate batch effect reduction, we measured normalized mutual information (NMI) and adjusted Rand index (ARI), both of which are commonly used to assess clustering performance.

Fig. \ref{fig:imputation_zero} displays UMAP plots of raw and imputed snRNA-seq data for the \textit{Leng} dataset.
The first column shows the UMAP of raw counts (before imputation), where cells of the same type are clustered together, but cells from different batches (patients) are noticeably separated.
This pattern suggests that the data is influenced more by technical factors than by biological characteristics.
When MAGIC is applied for imputation, clusters become scattered, and the batch effect remains uncorrected.
Furthermore, NMI and ARI values decrease compared to the raw counts, indicating poorer clustering performance.
Similarly, DCA fails to correct the batch effects, and it introduces an unintended connection between clusters of microglia and oligodendrocytes within the same patient group, implying inaccuracies in its imputed values.
In contrast, snRNA-seq data imputed with scHyena forms dense clusters of cells belonging to the same brain cell type.  
scHyena also achieves the highest NMI and ARI metric values, reflecting its strong performance in reducing batch effects.
Likewise, scMamba effectively reduces batch effects, as demonstrated by the UMAP visualization and the quantitative metrics.
These findings confirm that scMamba can impute snRNA-seq data with biologically meaningful values, thereby mitigating batch effects and enhancing data quality.

\subsection{scMamba improves robustness in DEG analysis.}
\begin{figure}[!t]
    \centering
    \includegraphics[width=0.99\linewidth]{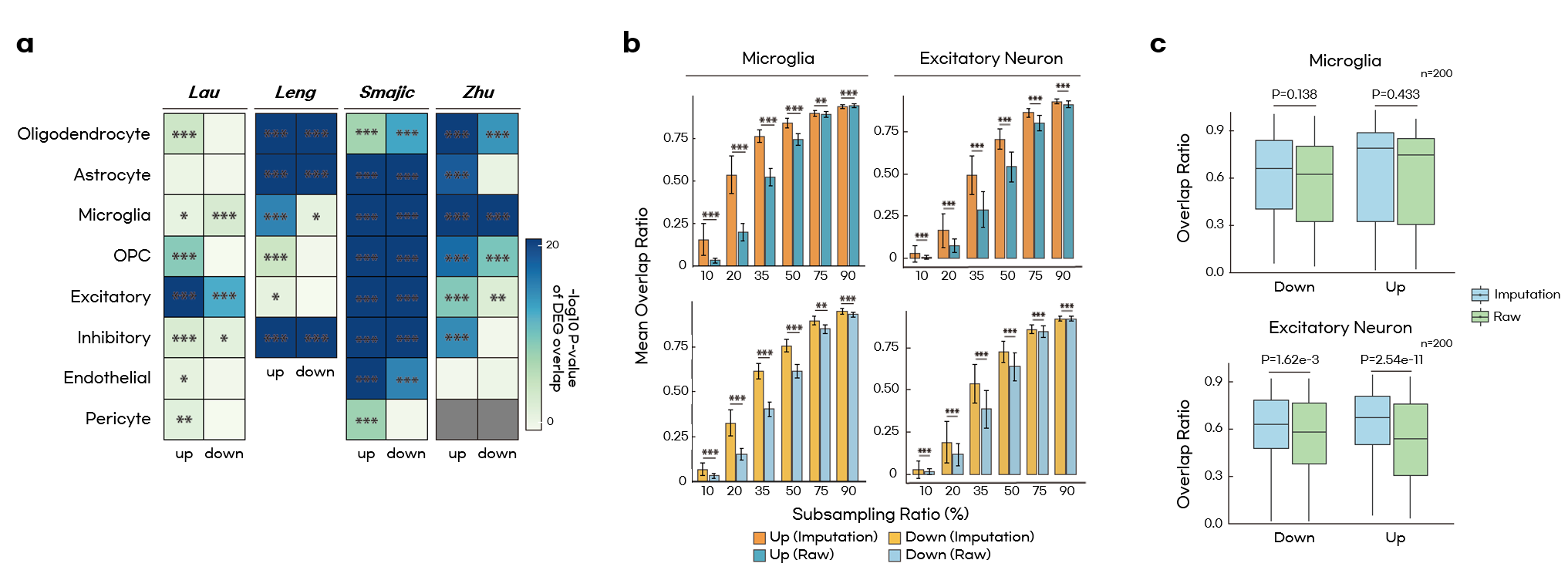}
    \caption{\bf\footnotesize
    Evaluating scMamba performance in single-cell differential expression analysis.
    \textbf{a}. The overlap ratio represents the intersection of DEGs between 50\% subsampled data and the original 100\% dataset.
    -log10 P-values were calculated using a paired t-test for each overlap ratio across 100 permutations.
    The grey box indicates missing values due to insufficient overlap to calculate P-values.
    The -log10 P-value range is capped at a maximum of 20 and a minimum of 0. Statistical significance is denoted as follows: *** for p < 0.001, ** for p < 0.01, and * for p < 0.05.
    \textbf{b}. Assessment of scMamba robustness through multiple subsampling ratios, showing the mean overlap ratio from 100 permutations for microglia (3717 cells) and excitatory neurons (980 cells) in the \textit{Smajic} dataset.
    Error bars indicate for standard deviation of each subsampled value.
    P-values were calculated using a paired t-test based on 100 permutations of the overlap ratio.
    Statistical significance is denoted as follows: *** for p < 0.001 and ** for p < 0.01.
    \textbf{c}. Boxplots showing the reproducibility of DEGs across all possible combinations(n=200) of half of the patients, compared to the original data, with and without imputation.
    Specifically, from a total of 5 patients and 6 neurotypical normal samples, we selected 3 patients and 3 neurotypical normal samples. Statistical significance was tested using a paired t-test.
    }
    \label{fig:deg}
\end{figure}

To assess the impact of scMamba on imputation, we focused on differential gene expression (DEG) analysis.
DEGs between diseased and neurotypical normal samples were identified with and without imputation, using data from two Alzheimer’s disease studies and two Parkinson’s disease studies, using MAST (v1.32.0).
Differential expression analysis was performed for eight major cell types using the FindMarkers pipeline in Seurat, with default parameters (adjusted P-value (FDR) < 0.05, log2FC > 0.25) in R.
The performance of scMamba imputation was evaluated by determining the fraction of overlapping DEGs between subsampled data and the original 100\% dataset across 100 permutations for each cell type in individual studies.
Subsampling was performed randomly in each round from the original dataset.
DEGs were categorized into upregulated and downregulated groups, which were then matched against the corresponding DEGs identified in each subsampled dataset.

In Fig. \ref{fig:deg}a, scMamba imputation with 50\% subsampling effectively recovered differentially expressed genes (DEGs) from the original dataset across most cell types and studies, outperforming the results obtained without imputation.
Despite variations across cell types and studies, both upregulated and downregulated genes showed statistically significant improvements in the recovery rate of DEGs after imputation.
To further investigate the impact of scMamba imputation on various subsampled ratios, we specifically focused on microglia and excitatory neurons from the \textit{Smajic} dataset (ref), selecting one large population and one smaller cell population, respectively.
Across varying subsampling ratios, scMamba imputation consistently retained more DEGs than the non-imputed data (Fig. \ref{fig:deg}b).
For example, with only 20\% subsampling, scMamba imputation recovered about 50\% of the original upregulated DEGs in microglia, while the non-imputed data recovered only 20\%.
Additionally, we compared the overlap ratio of DEGs by randomly selecting half of the samples from the same study, with and without imputation.
As expected, scMamba imputation enhanced reproducibility across all cell types, with a more significant impact in datasets with lower cell numbers (Fig. \ref{fig:deg}c and Supplementary Fig. \ref{fig:deg_supp}).

Overall, these results demonstrate that scMamba enhances both the quality and reproducibility of data, which is crucial for unbiased integration, especially when addressing the high variability introduced by low-quality postmortem brain tissue and the heterogeneity of disease.

\section*{Discussion}
In this work, we introduced scMamba, a pre-trained model designed to enhance the quality and utility of snRNA-seq analysis, particularly in studies of neurodegenerative diseases.
Built upon the Mamba model, scMamba can directly process raw snRNA-seq data without the need for dimensionality reduction.
We demonstrated that our model effectively learns meaningful representations of cells and genes through pre-training using masked expression modeling, even without incorporating additional information during pre-training.
As a result of these learned representations, scMamba achieved outstanding performance across various downstream tasks, including cell type classification, doublet detection, and snRNA-seq imputation.

In cell type classification, Seurat demonstrates lower performance than other methods at the subtype and subcluster levels.
As cell types are further subdivided, it becomes increasingly challenging to assign unique marker genes to each type.
Moreover, fine-grained clustering requires matching all clusters to specific types, a process that is not only time-consuming but also prone to reducing classification accuracy.

scBERT achieves strong classification performance but does not outperform the proposed method, possibly due to differences in data handling.
scBERT discretizes inputs by binning, which can lead to information loss.
In contrast, scMamba processes inputs directly without binning or discretization, allowing it to retain more information and outperform scBERT.

scHyena performs well in major cell type classification but exhibits diminished accuracy in subtype and subcluster classification, indicating a limited ability to distinguish finer cell types.
In contrast, scMamba consistently achieves superior performance across all classification levels.
We hypothesize that the ability of Mamba to selectively retain information based on inputs contributes to the superior performance of scMamba.

In the doublet detection experiments, DoubletDetection outperforms scMamba in precision and TNR, while scMamba demonstrates stronger performance across other metrics.
This indicates that scMamba identifies more doublets overall, including some additional false positives compared to DoubletDetection.
However, because thorough removal of doublets is critical, detecting doublets with high sensitivity is advantageous, even if it results in a slightly higher false positive rate, as long as other metrics remain robust.
Furthermore, we confirmed that scMamba surpasses DoubletDetection in precision and TNR at a fixed doublet identification rate.
From this perspective, scMamba offers a more effective approach than DoubletDetection for accurate doublet detection.

Although snRNA-seq is an advanced technology, analyzing the underlying causes of neurodegenerative disorders using snRNA-seq remains challenging because these datasets are typically derived from postmortem brain samples.
Variations in postmortem intervals can degrade sample quality, further complicating snRNA-seq analysis.
Additionally, disease heterogeneity poses significant obstacles to integrating snRNA-seq data from multiple sources.
To assess these challenges, accurate imputation of snRNA-seq data is crucial.
In our experiments, we demonstrated that scMamba accurately imputes snRNA-seq data, effectively correcting batch effects.
Furthermore, we found that imputation using scMamba enhances the robustness of differential gene expression (DEG) analysis.
These findings suggest that scMamba can facilitate the integration of snRNA-seq data from various sources, supporting large-scale studies.
Moreover, they highlight the potential scalability of scMamba for analyzing snRNA-seq data to uncover the causes of neurodegenerative disorders.

\section*{Methods}
\subsection{State space models.}
\begin{figure}[!t]
    \centering
    \includegraphics[width=0.6\linewidth]{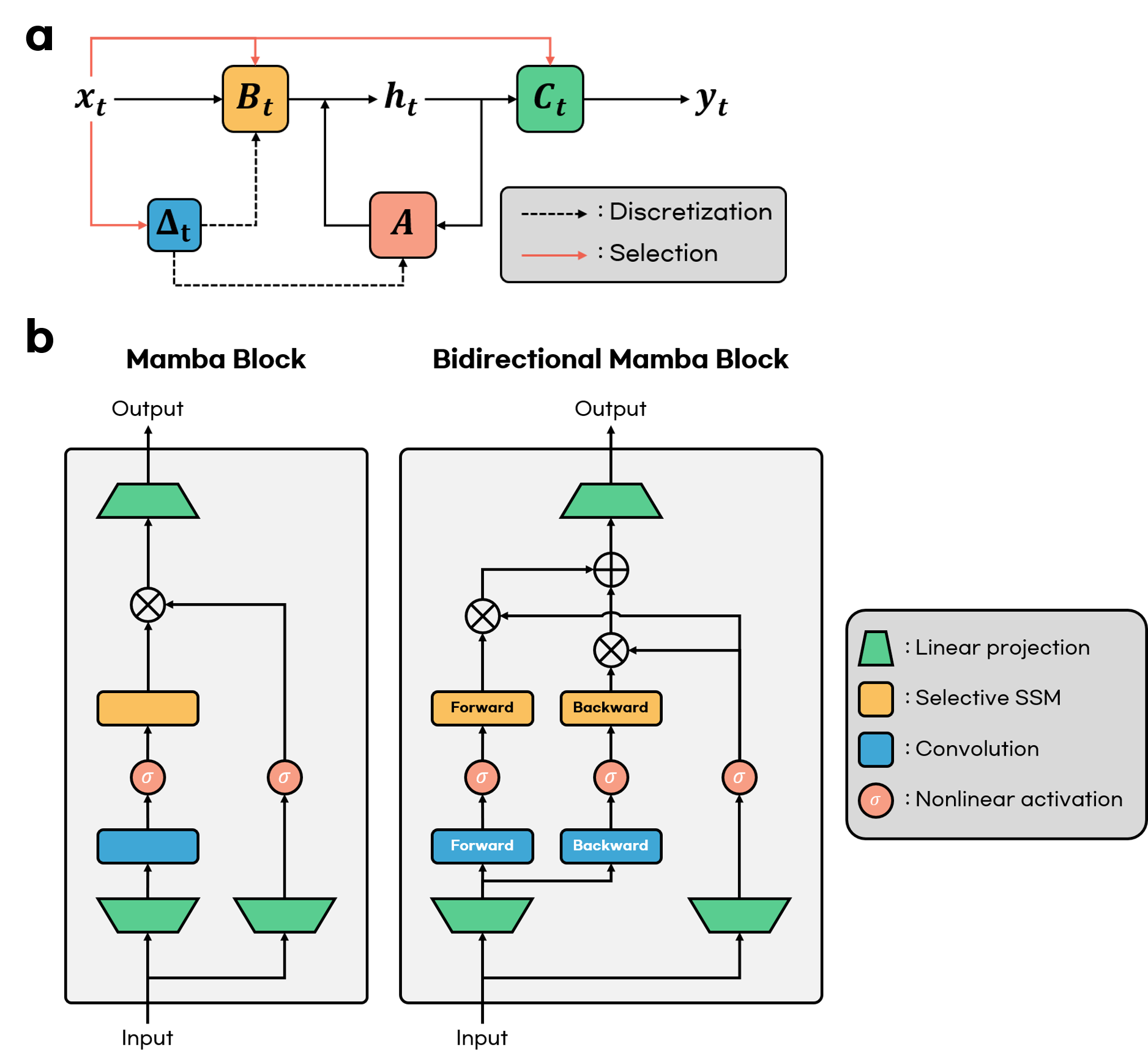}
    \caption{\bf\footnotesize
    State space models (SSMs) and Mamba.
    \textbf{a}. Selective SSMs dynamically adjust their parameters based on input data, enabling selective retention of information relevant to the input.
    It allows them to retain information based on input selectively.
    \textbf{b}. Selective state space models (SSMs) are organized into modular Mamba blocks, with the Mamba model being constructed by stacking multiple Mamba blocks.
    }
    \label{fig:mamba}
\end{figure}

State space models (SSMs) have been widely utilized in control theory.
Recently, SSMs have gained attention in deep learning for sequence data, as they can map input sequences to output sequences through a latent state.
In deep learning applications, where input data are typically discrete signals, the discretized form of SSMs is adopted.
Discrete SSMs can be expressed as follows:
\begin{equation}\label{eq:ssm}
    \begin{split}
        h_t &= \overline{\Ab} h_{t-1} + \overline{\Bb} x_t, \\
        y_t &= \Cb h_t,
    \end{split}
\end{equation}
where $\overline{\Ab}$, $\overline{\Bb}$, and $\Cb$ are the parameters of SSM.
Here, $\overline{\Ab}$ and $\overline{\Bb}$ are the discretized form of $\Ab$ and $\Bb$, defined as:
\begin{equation}
    \begin{split}
        \overline{\Ab} &= \exp{(\Delta \Ab)}, \\
        \overline{\Bb} &= (\Delta \Ab)^{-1} (\exp{(\Delta \Ab) - \Ib} \cdot \Delta \Bb,
    \end{split}
\end{equation}
where $\Delta$ represents the learnable step size parameter.

While the self-attention mechanism in Transformer \cite{vaswani2017attention} explicitly calculates relationships between inputs, SSMs implicitly capture these relationships through compressed state variables.
This allows SSMs to handle longer sequences with significantly lower computational complexity than Transformers.
However, unlike self-attention maps, which are dynamically computed based on the input, the parameters of SSMs remain static and do not adapt to the input.
This lack of input-dependent adaptability limits the performance of SSMs in certain scenarios.

The selective SSM (S6) \cite{gu2024mamba} was introduced to address the limitation of standard SSMs.
As illustrated in Fig. \ref{fig:mamba}a, S6 allows $\Bb$, $\Cb$, and $\Delta$ to become input-dependent as follows:
\begin{equation}\notag
    \begin{split}
        \Bb &= \textbf{Linear}_{N}(x), \\
        \Cb &= \textbf{Linear}_{N}(x), \\
        \Delta &= \log{(1 + \exp{(\textbf{Parameter} + \textbf{Broadcast}_D(\textbf{Linear}_1(x)))})}.
    \end{split}
\end{equation}
This adaptation enables SSMs to be content-aware, allowing them to selectively retain or discard information based on input.

\subsection{Mamba.}
Selective SSMs can be organized into modular blocks.
Recently, the Mamba block \cite{gu2024mamba} was introduced, which utilizes selective SSMs, as shown in Fig. \ref{fig:mamba}b.
By stacking multiple Mamba blocks, a structure resembling the Transformer decoder can be constructed.

However, since SSMs are causal systems, they are inherently limited in their application to non-temporal data, such as visual data.
To address this limitation, some studies propose bidirectional Mamba blocks \cite{liu2024vmamba,zhu2024vision}, and Fig. \ref{fig:mamba}b illustrates an example of a bidirectional Mamba block.

\subsection{scMamba.}
The proposed scMamba model is illustrated in Fig. \ref{fig:scmamba}a.
To effectively process long snRNA-seq sequences, we adopted the Mamba architecture composed of multiple Mamba blocks.
In the context of snRNA-seq data, the concept of time causality does not apply, as any gene can potentially relate to others regardless of its position in the sequence.
To account for this, we implemented a non-causal bidirectional Mamba block, enabling the output to depend on the entire input across all sequence positions.

An input snRNA-seq consists of the normalized expression levels of $L$ genes, denoted as ($C_1, C_2, \dots, C_L$).
Unlike natural language, where words are tokenized into discrete tokens and each token is mapped to a unique embedding, gene expression levels are continuous values that cannot be directly discretized.
A previous method \cite{yang2022scbert} addressed this issue by discretizing expression values into bins.
However, this approach risks significant information loss in the snRNA-seq data.
To mitigate this concern, we encode the expression levels into expression embeddings ($E_{C_1}, E_{C_2}, \dots, E_{C_L}$) using a linear adapter layer.
This approach preserves the continuous nature of gene expression data, ensuring no loss of information during the encoding process.

Another key difference between language and snRNA-seq data is that the order of genes in snRNA-seq has no inherent meaning.
Instead, it is crucial to provide information indicating which gene's expression level corresponds to each position in the sequence.
To achieve this, we incorporate gene embeddings into the scMamba model, replacing the positional encodings used in traditional Transformers.
In this approach, each gene is assigned a unique embedding ($E_{G_1}, E_{G_2}, \dots, E_{G_L}$), which is then added to the expression embeddings.
This strategy ensures that the scMamba model receives explicit gene-related information, enabling it to effectively interpret the snRNA-seq data.
After combining the gene embeddings with the expression embeddings, the final input embeddings ($E_1, E_2, \dots, E_L$) are generated and fed into the Mamba blocks.

\subsection{Pre-training.}
Fig. \ref{fig:scmamba}a illustrates the pre-training process of the scMamba model.
To pre-train our model, we employ a technique called masked expression modeling (MEM), inspired by the concept of masked language modeling \cite{devlin2018bert}.
In this approach, a subset of input embeddings is randomly replaced with an [MASK] embedding, and scMamba is trained to predict the expression levels of the masked genes.
The masking probability is set to 0.15, and only nonzero values are masked, as distinguishing between true and false zero values is not feasible.

Since genes can interact regardless of their positions, predicting masked expressions requires considering relationships between all genes.
To achieve this, scMamba uses bidirectional Mamba blocks, enabling it to predict expression levels at masked positions.
The pre-training objective is defined as:
\begin{equation}\label{eq:mem}
    \ell_{MEM} = \sum_{i \in M}(C_{i} - \hat{C}_{i})^2,
\end{equation}
where $M$ denotes the set of masked indices, and $C_i$ and $C'_{i}$ represent the true and predicted gene expression levels, respectively.
This pre-training process allows scMamba to learn generalizable features of both cells and genes.

\subsection{Cell type classification and doublet detection.}
Cell type classification and doublet detection are among the most important tasks in snRNA-seq analysis.
Fig. \ref{fig:scmamba}b illustrates the fine-tuning process of scMamba for these tasks.
To adapt the pre-trained scMamba model for cell type classification or doublet detection, we insert [CLS] embeddings into the input data embeddings.
While the [CLS] token or embedding is typically prepended to the input, we observed that placing three [CLS] embeddings—at the beginning, middle, and end of the input—enhanced performance.

After the input embeddings pass through the Mamba blocks, the three [CLS] embeddings are concatenated and fed into a classification head.
The output of the classification head generates logits, and scMamba is fine-tuned using the cross-entropy loss, $\ell_{cls} = -\sum_{i = 1}^{N_c}y_{i}\log{p_i}$, where $N_c$ is the number of classes, $y_i$ is the true class label, and $p_i$ is the Softmax probability derived from the output of scMamba.
The hyperparameters for pre-training and fine-tuning can be found in Supplementary Table \ref{tab:hyper_supp}.

\subsection{snRNA-seq imputation.}
Imputing missing values in snRNA-seq data is essential due to the high prevalence of zeroes resulting from dropout events.
One strategy for imputation involves directly adapting the pre-training approach.
However, in pre-training, zero values are not masked, which may lead the model to learn to replace true zero values with other values incorrectly.
Alternatively, if zero values are masked at the same probability as non-zero values, the model may disproportionately predict zeroes, given that most values in snRNA-seq data are zeroes.

To mitigate this issue, we apply different masking probabilities for zero and non-zero values.
Specifically, non-zero values are masked with a probability of 0.4, while zero values are masked with a lower probability of 0.04.
This adjustment balances the masking process for zero and non-zero values.

Despite this improvement, a challenge persists in distinguishing true zero values from false zeroes caused by dropout.
In some cases, the model may impute false zeroes as true zeroes.
Fortunately, our scMamba model leverages gene embeddings, enabling it to learn and incorporate the expression level tendencies of individual genes.

Fig. \ref{fig:scmamba}c depicts the fine-tuning process for snRNA-seq imputation.
The loss function for the imputation task remains consistent with the pre-training loss function, as shown in Equation \eqref{eq:mem}.
To ensure high-quality training data and improve model performance, we only used data with a UMI count greater than 4,000 for fine-tuning.

\subsection{Data preparation.}
\paragraph{Collection of published human neurodegenerative brain snRNA-seq data.}
For this study, 14 distinct published snRNA-seq processed datasets were collected as gene-by-cell count matrices, with the following identifiers: GSE140231 \cite{agarwal2020single}, GSE148822 \cite{gerrits2021distinct}, GSE178265 \cite{kamath2022single}, GSE157827 \cite{lau2020single}, GSE147528 \cite{leng2021molecular}, GSE129308 \cite{otero2022molecular}, GSE174367 \cite{morabito2021single}, GSE167494 \cite{sadick2022astrocytes}, GSE157783 \cite{smajic2022single}, GSE160936 \cite{smith2022diverse}, GSE184950 \cite{wang2022single}, GSE163577 \cite{yang2022human}, GSE188545 \cite{zhang2023single}, and GSE202210 \cite{zhu2024single}.
In addition, we utilized a custom dataset containing approximately 500,000 cells, referred to as the \textit{Jung} dataset.

To integrate the data based on consistent transcriptomic information, Ensembl stable gene IDs were used in place of gene symbols.
All datasets were concatenated using the union of Ensembl IDs, resulting in 61,325 genes.
Undetected genes in each cell were assigned a value of 0.
Genes located on chromosome Y or lacking annotation in the GRCh38 version 108 GTF file were filtered out.
Additionally, only genes detected in at least 0.5\% of all cells—excluding the \textit{Kamath} dataset—were retained, yielding 19,306 unique Ensembl IDs.

Among the data, \textit{Lau} \cite{lau2020single}, \textit{Leng} \cite{leng2021molecular}, \textit{Smajic} \cite{smajic2022single}, \textit{Zhu} \cite{zhu2024single}, and \textit{Jung} datasets were selected for downstream task, while the remaining 10 datasets were used for pre-training scMamba.
The pre-training dataset comprises approximately 1.6 million cells.
For Alzheimer’s disease, \textit{Lau} and \textit{Leng} datasets were chosen, representing three distinct brain regions-the prefrontal cortex, caudal entorhinal cortex, and superior frontal gyrus-while preserving the large size of the nucleus.
For Parkinson’s disease, \textit{Smajic} and \textit{Zhu} datasets were selected, covering the substantia nigra and frontal cortex, respectively.
The \textit{Jung} dataset includes cases of both Alzheimer’s disease and Parkinson’s disease and spans regions such as the prefrontal cortex, hippocampus, and substantia nigra.
The structure and detailed information about the datasets are provided in Supplementary Table \ref{tab:data_celltype_supp} and \ref{tab:data_disease_supp}.

\paragraph{Cell type annotation based on unsupervised clustering.}
The collected count matrices were preprocessed following the canonical SCANPY analysis pipeline \cite{wolf2018scanpy}.
The resulting data were integrated with the public datasets based on a unified list of 19,306 genes.
Patients with less than 200 cells were filtered out, leaving a total of 2,408,023 nuclei from 461 patients.
To identify doublets arising from experimental artifacts in the single-cell technique, we applied Scrublet \cite{wolock2019scrublet} to calculate doublet scores and predict doublets for each single cell.
These doublet scores were used to annotate doublet-enriched clusters.
Cell clustering into distinct cell types was performed using the top 2,000 highly variable genes (HVGs), selected based on analytic Pearson residuals \cite{lause2021analytic}.
These genes were utilized for principal component analysis (PCA) and clustering.
Total UMI counts per cell were normalized to 50,000, log2-transformed with a pseudo-count of 1, and scaled using scanpy.pp.scale.
PCA coordinates were then computed using scanpy.pp.pca with default parameters.
To mitigate batch effects across patients, Harmony correction \cite{korsunsky2019fast} was applied to the PCA coordinates.
Neighborhoods for every single cell were calculated using scanpy.pp.neighbors with the parameter of 20 PCA components and 40 nearest neighbors.
The data were further reduced to a two-dimensional plane using UMAP for visualization.
Finally, Leiden clustering was performed at a resolution of 1.8, resulting in 69 distinct clusters.

To annotate cell types for each cluster, the expression levels of known marker genes for major human brain cell types were analyzed across the clusters.
Using a panel of 40 distinct marker genes, 51 clusters were assigned to 11 cell types, including 8 major brain cell types: Oligodendrocytes (CLDN11, MBP), Astrocytes (AQP4, ALDH1L1), Microglia (C1QC, CSF1R), Endothelial cells (CLDN5, FLT1), Pericytes (PDGFRB), Oligodendrocyte progenitor cells (OPCs; PDGFRA, VCAN), Excitatory neurons (SYT1, SLC17A7, SLC17A6), and Inhibitory neurons (SYT1, GAD1, GAD2).
Additionally, three subtypes were identified: Neuron subtype 1 (SYT1, SLC17A6, GAD2), Neuron subtype 2 (SYT1, but neither SLC17A6 nor GAD2), and Myeloid subtype 1 (GNLY, CD44).

One cluster, which did not exhibit a clear expression pattern for any marker genes, was labeled as ``unidentified''.
Clusters with an average doublet score exceeding 0.1 were classified as doublets.
For the cell type classification task, only single cells annotated to one of the 8 major cell types were considered.

\paragraph{Preprocessing.}
As part of our preprocessing pipeline for snRNA-seq data, we first filtered out cells with a total gene expression level below 200.
Next, we normalized the gene expression values by scaling the total expression count of each cell to 10,000.
Finally, log normalization ($\log(x + 1)$) was applied to generate the final preprocessed dataset.

\subsection{Single-Cell Differential Expression Analysis.}
We identified DEGs from imputed and raw data across four studies using MAST (v1.32.0), a method specifically designed for single-cell data analysis.
Differential expression analysis was performed for eight major cell types, comparing disease and normal cells, following the FindMarkers pipeline in Seurat with default parameters (adjusted P-value (FDR) < 0.05, log2FC > 0.25) in R.
The efficiency of scMamba was evaluated by determining the fraction of overlapping DEGs between 50\% subsampled data and the original 100\% dataset across 100 permutations.
Subsampling was performed randomly each round from the original data.
DEGs from the original data were categorized into upregulated and downregulated groups, which were then matched against upregulated and downregulated DEGs identified in each 50\% subsampled dataset.
The overlap ratio was calculated as the intersection of DEGs between the subsampled and original data, divided by the total DEGs from the original data.
The matching DEGs between subsampled and original datasets was summarized by calculating the mean overlap ratio.
P-values were determined using a paired t-test and transformed into -log10 P-values for easier visualization of significance.
The mean overlap ratio of microglia and excitatory neurons presented in Fig. \ref{fig:deg}b represents the average of overlapping DEGs across 100 permutations at each subsampling ratio (10\%, 20\%, 35\%, 50\%, 75\%, 90\%).
To evaluate the reproducibility between patients, we randomly selected subsets of samples, specifically 3 patient samples and 3 normal samples, from a total of 5 patient samples and 6 normal samples covering all possible combinations (n=200).
Differential expression analysis was then conducted on subsets of the eight major cell types.
The overlap ratio of DEGs was compared within datasets, where ``with imputation'' refers to comparisons between imputed data, and ``without imputation" refers to comparisons between raw data.
The significance of these ratios was assessed using a paired t-test.



\section*{References}
\bibliographystyle{naturemag}
\bibliography{ref}

 \begin{addendum}
{\color{black} \item[Correspondence] Correspondence to Jong Chul Ye or Inkyung Jung.}
 \item  This research was supported by National Research Foundation of Korea(NRF) (**RS-2023-00262527**).
{
\item[Author Contributions] These authors contributed equally: Gyutaek Oh, Baekgyu Choi.\\
G.O. developed the code, conducted experiments, analyzed the results, and drafted and revised the manuscript.
B.C. collected the data, analyzed the results, and drafted and revised the manuscript.
S.J. analyzed the results and revised the manuscript.
J.C.Y. and I.J. supervised the project, guided its conceptualization and discussions, and prepared the manuscript.}
 \item[Competing Interests] 
The Authors declare no competing interests.

\end{addendum}

\newpage

\section*{Supplementary Information}
\setcounter{figure}{0}
\renewcommand{\figurename}{Supplementary Fig.}
\renewcommand{\thefigure}{\arabic{figure}}
\setcounter{table}{0}
\renewcommand{\tablename}{Supplementary Table}
\renewcommand{\thetable}{\arabic{table}}

\begin{table}[!h]
\caption{Hyperparameters for pre-training and fine-tuning.
}
\label{tab:hyper_supp}
\begin{center}
\resizebox{0.8\linewidth}{!}{
\begin{tabular} {c|cc}
\hline
\ Task  & Pre-Training    & Fine-Tuning \\
\hline
\ Learning Rate & 1e$^{-3}$  & 1e$^{-4}$ \\
\ Training Epoch & 2  & 5, 10 \\
\hline
\ Optimizer & \multicolumn{2}{c}{AdamW} \\
\ Optimizer Momentum    & \multicolumn{2}{c}{$\beta_1, \beta_2 = 0.9, 0.95$} \\
\ Weight Decay  & \multicolumn{2}{c}{0.1} \\
\ Learning Rate Scheduler   & \multicolumn{2}{c}{Cosine decay} \\
\ Batch Size    & \multicolumn{2}{c}{16} \\
\hline
\end{tabular}
}
\end{center}
\end{table}

\begin{sidewaystable}
\caption{The distribution of cell types in the datasets (AC: astrocyte, MG: microglia, OL: oligodendrocyte, OPC: oligodendrocyte progenitor cell, EXN: excitatory neuron, INN: inhibitory neuron, EC: endothelial cell, PC: pericyte, DT: doublet, ETC: others).
}
\label{tab:data_celltype_supp}
\begin{center}
\resizebox{0.99\linewidth}{!}{
\begin{tabular} {c|c|cccccccccc|c}
\hline
\ Task  & Dataset  & AC  & MG    & OL    & OPC   & EXN   & INN   & EC    & PC    & DT    & ETC   & Total (Train/Validation/Test) \\
\hline\hline
\multirow{10}{*}{Pre-Training}    & \textit{Agarwal}  & 415    & 273    & 3,933   & 413    & 6,557    & 2,734    & 14    & 10  & 1,703     & 588    & 16,640 (16,640/0/0)   \\
\   & \textit{Gerrits}  & 113,506    & 127,022    & 35,747   & 679    & 4,070    & 2,854    & 20,495    & 7,965  & 37,107     & 28,154    & 377,599 (377,599/0/0)   \\
\   & \textit{Kamath}  & 40,848    & 34,816    & 185,451   & 15,148    & 28,031    & 11,570    & 5,786    & 3,927  & 5,603     & 99,132    & 430,312 (430,312/0/0)   \\
\   & \textit{Marcos}  & 385    & 11    & 169   & 69    & 82,452    & 13,782    & 25    & 13  & 66     & 4,175    & 101,147 (101,147/0/0)   \\
\   & \textit{Morabito}  & 4,253    & 3,600    & 36,551   & 2,567    & 5,775    & 5,524    & 126    & 174  & 1,179     & 269    & 60,018 (60,018/0/0)   \\
\   & \textit{Sadick}  & 52,299    & 3,955    & 35,231   & 3,148    & 13,163    & 6,405    & 512    & 1,470  & 4,599     & 3,869    & 124,651 (124,651/0/0)   \\
\   & \textit{Smith}  & 55,351    & 25,766    & 388   & 1,230    & 1,208    & 2,390    & 924    & 451  & 3,504     & 2,173    & 93,385 (93,385/0/0)   \\
\   & \textit{Wang}     & 5,942     & 7,803     & 81,378    & 8,361     & 5,987     & 840       & 4,804    & 2,137  & 21,202    & 6,716     & 145,170 (145,170/0)    \\
\   & \textit{Yang}  & 22,295    & 2,190    & 30,372   & 2,477    & 720    & 1,693    & 38,293    & 23,955  & 38,176     & 3,043    & 163,214 (163,214/0/0)   \\
\   & \textit{Zhang}  & 6,039    & 3,807    & 22,245   & 4,837    & 20,429    & 10,328    & 171    & 185  & 4,036     & 723    & 72,800 (72,800/0/0)   \\
\hline
\multirow{5}{*}{Downstream Tasks}   & \textit{Lau}  & 12,157    & 4,719     & 30,571    & 9,223     & 50,572    & 18,932    & 540      & 444    & 12,427    & -         & 139,585 (96,311/12,522/30,752) \\
\   & \textit{Leng}     & 6,650     & 2,260     & 11,904    & 3,038     & 19,926    & 9,656     & 194      & 135    & 3,590     & -         & 57,353 (39,911/5,059/12,383) \\
\   & \textit{Smajic}   & 5,018     & 3,717     & 20,956    & 2,674     & 980       & 705       & 1,641    & 673    & 1,479     & -         & 37,843 (26,929/3,692/7,222)  \\
\   & \textit{Zhu}      & 7,077     & 4,386     & 22,773    & 4,737     & 19,200    & 11,763    & 233      & 156    & 3,425     & -         & 73,750 (52,753/6,835/14,162)  \\
\   & \textit{Jung}     & 36,605     & 16,860     & 144,237    & 17,993     & 121,246    & 43,161    & 9,897      & 6,463    & 54,164     & -         & 450,626 (323,754/36,095/90,777)  \\
\hline
\end{tabular}
}
\end{center}
\end{sidewaystable}

\begin{table}
\caption{The distribution of disease in the datasets (CN: cognitively normal, AD: Alzheimer's disease, PD: Parkinson's disease, LB: Lewy body dementia, ETC: unknown).
}
\label{tab:data_disease_supp}
\begin{center}
\resizebox{0.85\linewidth}{!}{
\begin{tabular} {c|c|ccccc|c}
\hline
\ Task  & Dataset  & CN  & AD    & PD    & LB    & ETC   & Total (Train/Validation/Test) \\
\hline\hline
\multirow{10}{*}{Pre-Training}    & \textit{Agarwal}  & 12    & 0    & 0   & 0    & 0    & 12 (12/0/0)   \\
\   & \textit{Gerrits}  & 16    & 20    & 0   & 0    & 0    & 36 (36/0/0)   \\
\   & \textit{Kamath}  & 44    & 0    & 32   & 18    & 3    & 97 (97/0/0)   \\
\   & \textit{Marcos}  & 8    & 19    & 0   & 0    & 0    & 27 (27/0/0)   \\
\   & \textit{Morabito}  & 7    & 11    & 0   & 0    & 0    & 18 (18/0/0)   \\
\   & \textit{Sadick}  & 10    & 14    & 0   & 0    & 0    & 24 (24/0/0)   \\
\   & \textit{Smith}  & 12    & 12    & 0   & 0    & 0    & 24 (24/0/0)   \\
\   & \textit{Wang}     & 9     & 0     & 19    & 0     & 5     & 33 (33/0)    \\
\   & \textit{Yang}  & 12    & 13    & 0   & 0    & 0    & 25 (25/0/0)   \\
\   & \textit{Zhang}  & 6    & 6    & 0   & 0    & 0    & 12 (12/0/0)   \\
\hline
\multirow{5}{*}{Downstream Tasks}   & \textit{Lau}  & 9    & 12     & 0    & 0     & 0     & 21 (15/2/4) \\
\   & \textit{Leng}     & 6     & 14     & 0    & 0     & 0     & 20 (14/2/4) \\
\   & \textit{Smajic}   & 6     & 0     & 5    & 0     & 0     & 11 (7/2/2)  \\
\   & \textit{Zhu}      & 6     & 0     & 6    & 0     & 0     & 12 (8/2/2)  \\
\   & \textit{Jung}     & 9     & 8     & 5    & 3     & 0     & 25 (15/4/6)  \\
\hline
\end{tabular}
}
\end{center}
\end{table}

\begin{sidewaystable}[!h]
\caption{\bf\footnotesize
Average F1 scores for cell type, subtype, and subcluster classification of various methods.
}
\label{tab:celltype_supp}
\begin{center}
\resizebox{0.8\linewidth}{!}{
\begin{tabular} {c|c|ccc|ccc|ccc}
\hline
\multirow{2}{*}{Data}  & \multirow{2}{*}{Method}    & \multicolumn{3}{c|}{F1 Score (Cell Type)}  & \multicolumn{3}{c|}{F1 Score (Subtype)}  & \multicolumn{3}{c}{F1 Score (Subcluster)} \\
\cline{3-11}
\   &   & Macro    & Micro    & Weighted    & Macro    & Micro    & Weighted    & Macro    & Micro    & Weighted \\
\hline\hline
\multirow{5}{*}{\textit{Lau}}    & Seurat  & 0.9938 & 0.9953 & 0.9953   & 0.4162 & 0.6305 & 0.6219   & -    & -    & - \\
\   & SciBet     & 0.9943 & 0.9968 & 0.9968   & \underline{0.6557} & 0.7437 & 0.7453   & \underline{0.5206}    & 0.6486    & 0.6497 \\
\   & scBERT     & 0.9939 & 0.9978 & 0.9978   & 0.6457 & \underline{0.7895} & \underline{0.7838}   & 0.5034    & \underline{0.6975}    & \underline{0.6907} \\
\   & scHyena    & \underline{0.9960} & \textbf{0.9982} & \textbf{0.9982}   & 0.5352 & 0.7800 & 0.7696   & 0.3983    & 0.6579    & 0.6430 \\
\   & Proposed   & \textbf{0.9983} & \textbf{0.9982} & \textbf{0.9982}   & \textbf{0.6916} & \textbf{0.8168} & \textbf{0.8137}   & \textbf{0.5575} & \textbf{0.7425} & \textbf{0.7402} \\
\hline
\multirow{5}{*}{\textit{Leng}}    & Seurat  & 0.8248 & 0.9936 & 0.9926   & 0.4088 & 0.5649 & 0.5548   & -    & -    & - \\
\   & SciBet     & \underline{0.9905} & 0.9965 & 0.9965   & \underline{0.6194} & 0.7069 & 0.6969   & \underline{0.4843}    & 0.5791    & 0.5780 \\
\   & scBERT     & 0.9800 & 0.9961 & 0.9960   & 0.6027 & \textbf{0.7703} & \underline{0.7588}   & 0.4549    & \underline{0.6638}    & \underline{0.6456} \\
\   & scHyena    & 0.9796 & \underline{0.9970} & \underline{0.9970}   & 0.4131 & 0.6855 & 0.6615   & 0.2972    & 0.5921    & 0.5590 \\
\   & Proposed   & \textbf{0.9947} & \textbf{0.9976} & \textbf{0.9976}   & \textbf{0.6503} & \underline{0.7641} & \textbf{0.7595}   & \textbf{0.5203} & \textbf{0.6669} & \textbf{0.6610} \\
\hline
\multirow{5}{*}{\textit{Smajic}}    & Seurat  & 0.9426 & 0.9922 & 0.9920   & 0.3016 & 0.5413 & 0.5351   & -    & -    & - \\
\   & SciBet     & 0.9481 & 0.9946 & 0.9941   & \textbf{0.4738} & 0.6888 & 0.6934   & \underline{0.3557}    & 0.6068    & 0.6040 \\
\   & scBERT     & 0.9532 & 0.9950 & 0.9946   & 0.2920 & 0.7383 & 0.7052   & 0.1652    & 0.5807    & 0.5399 \\
\   & scHyena    & \textbf{0.9843} & \textbf{0.9979} & \textbf{0.9979}   & 0.2953 & \underline{0.7549} & \underline{0.7384}   & 0.2145    & \underline{0.6702}    & \underline{0.6498} \\
\   & Proposed   & \underline{0.9742} & \underline{0.9967} & \underline{0.9968}   & \underline{0.3807} & \textbf{0.8047} & \textbf{0.7948}   & \textbf{0.3739} & \textbf{0.7396} & \textbf{0.7395} \\
\hline
\multirow{5}{*}{\textit{Zhu}}    & Seurat  & \underline{0.9921} & 0.9962 & 0.9962   & 0.5374 & 0.6289 & 0.6265   & -    & -    & - \\
\   & SciBet     & 0.9892 & 0.9943 & 0.9943   & \underline{0.6805} & 0.6160 & 0.6389   & \underline{0.4881}    & 0.5153    & 0.5351 \\
\   & scBERT     & 0.9529 & 0.9938 & 0.9937   & 0.5703 & \underline{0.7406} & \underline{0.7139}   & 0.3606    & 0.5754    & 0.5442 \\
\   & scHyena    & 0.9880 & \underline{0.9968} & \underline{0.9968}   & 0.5183 & 0.7280 & 0.7103   & 0.3439    & \underline{0.6269}    & \underline{0.6016} \\
\   & Proposed   & \textbf{0.9940} & \textbf{0.9974} & \textbf{0.9974}   & \textbf{0.6889} & \textbf{0.7770} & \textbf{0.7775}   & \textbf{0.5667} & \textbf{0.7025} & \textbf{0.7008} \\
\hline
\multirow{5}{*}{\textit{Jung}}    & Seurat  & 0.9245 & 0.9209 & 0.9227   & 0.3002 & 0.4478 & 0.4646   & -    & -    & - \\
\   & SciBet     & 0.9455 & 0.9452 & 0.9467   & 0.6466 & 0.5948 & 0.5968   & 0.4898    & 0.4805    & 0.4896 \\
\   & scBERT     & 0.9747 & \textbf{0.9791} & \textbf{0.9791}   & \underline{0.6900} & \underline{0.7028} & \underline{0.7038}   & \underline{0.5205}    & \underline{0.6161}    & \underline{0.6069} \\
\   & scHyena    & \underline{0.9751} & \underline{0.9790} & \underline{0.9790}   & 0.6142 & 0.6840 & 0.6752   & 0.4765    & 0.6111    & 0.5938 \\
\   & Proposed   & \textbf{0.9772} & 0.9766 & 0.9768   & \textbf{0.7135} & \textbf{0.7168} & \textbf{0.7194}   & \textbf{0.5911} & \textbf{0.6587} & \textbf{0.6578} \\
\hline
\end{tabular}
}
\end{center}
\end{sidewaystable}

\begin{sidewaysfigure}[!h]
    \centering
    \includegraphics[width=0.99\linewidth]{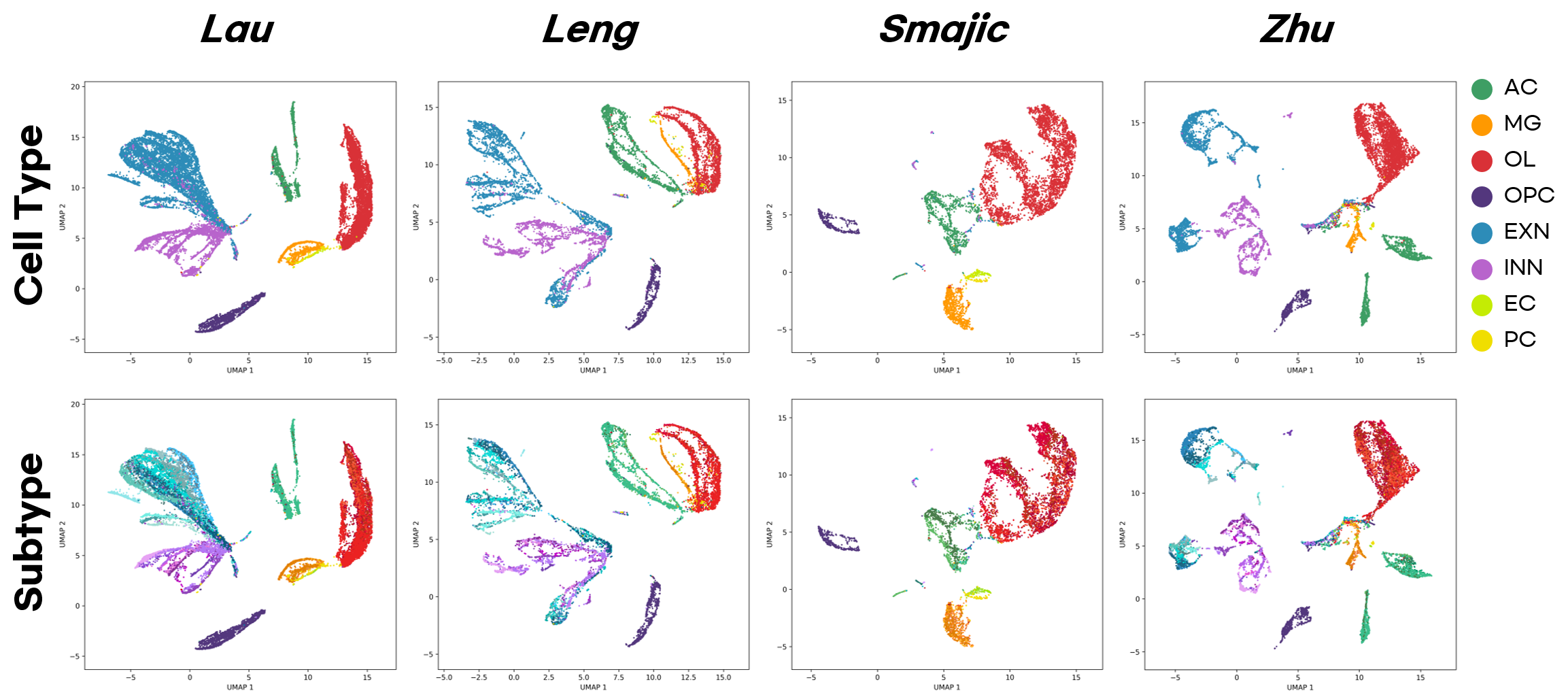}
    \caption{\bf\footnotesize
    UMAP visualization of cell embeddings from the pre-trained scMamba model.
    Each UMAP is colored based on 8 major cell types or 72 subtypes.
    (AC: astrocyte, MG: microglia, OL: oligodendrocyte, OPC: oligodendrocyte progenitor cell, EXN: excitatory neuron, INN: inhibitory neuron, EC: endothelial cell, PC: pericyte).
    }
    \label{fig:embeddings_supp}
\end{sidewaysfigure}

\begin{sidewaysfigure}[!h]
    \centering
    \includegraphics[width=0.85\linewidth]{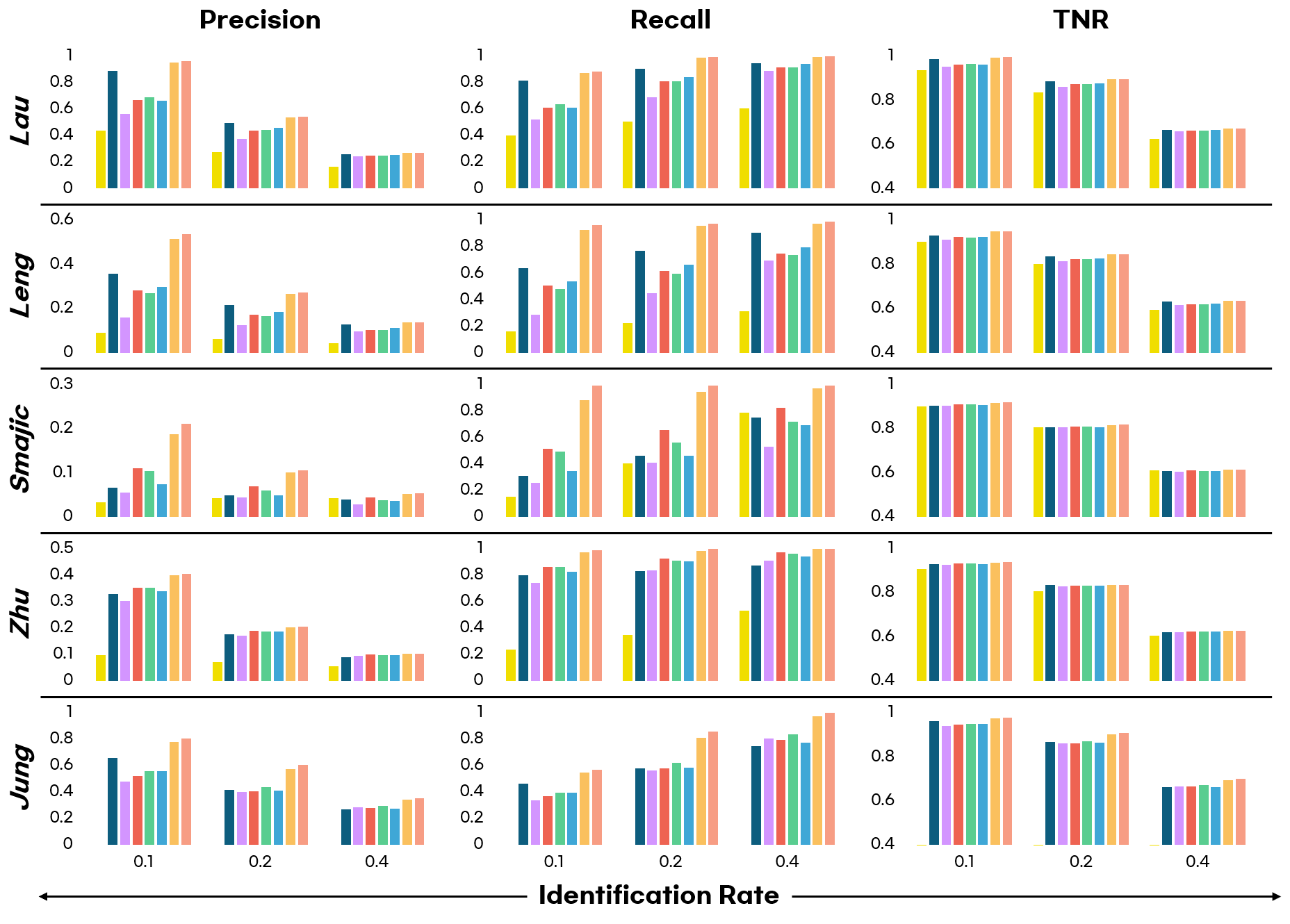}
    \caption{\bf\footnotesize
    Precision, recall, and TNR for in vivo doublet detection by each method across datasets.
    The x-axis of the histograms represents the identification rate.
    }
    \label{fig:doublet_invivo_supp}
\end{sidewaysfigure}

\begin{sidewaysfigure}[!h]
    \centering
    \includegraphics[width=0.85\linewidth]{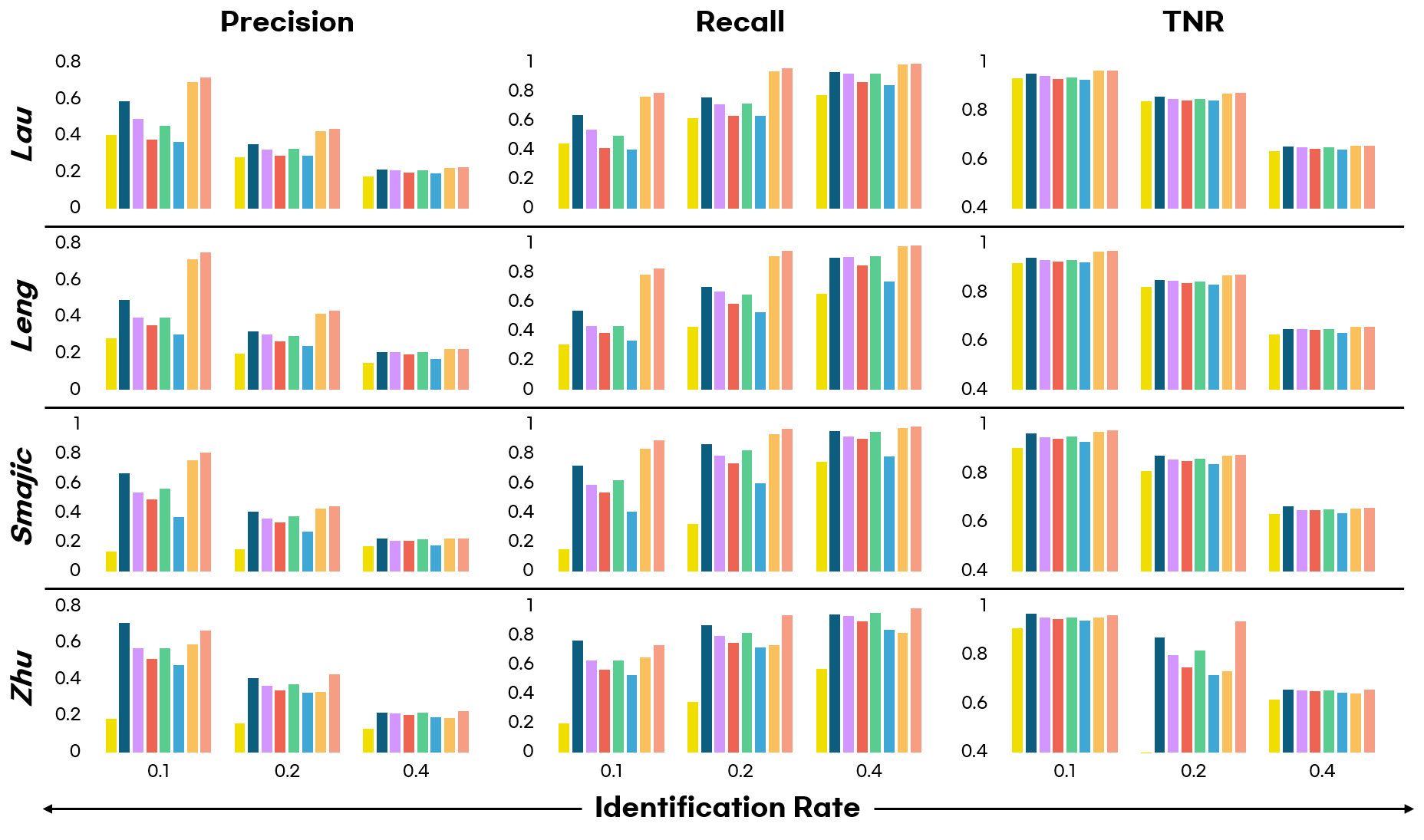}
    \caption{\bf\footnotesize
    Precision, recall, and TNR for simulated doublet detection by each method across datasets.
    The x-axis of the histograms represents the identification rate.
    }
    \label{fig:doublet_simul_supp}
\end{sidewaysfigure}

\begin{figure}[!h]
    \centering
    \includegraphics[width=0.99\linewidth]{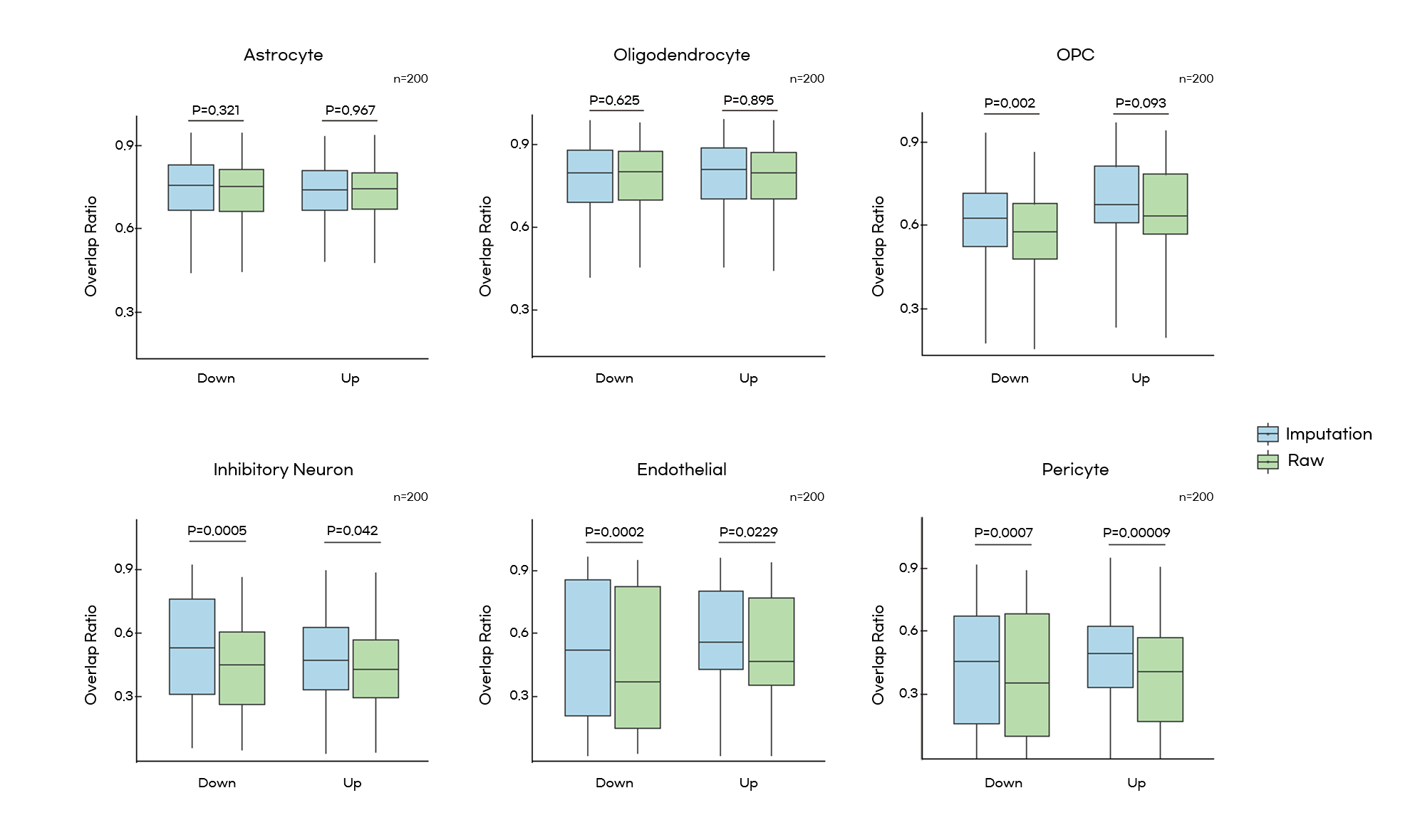}
    \caption{\bf\footnotesize
    Boxplots showing for the fraction of DEG overlap for six cell types after selecting half of samples within the same study.
    The number of cells for each cell type is shown together: Astrocytes = 5,018, Oligodendrocytes = 20,956, OPCs = 2,674, Inhibitory neurons = 705, Endothelial cells = 1,641, and Pericytes = 673.
    P-values were calculated using paired t-test.
    }
    \label{fig:deg_supp}
\end{figure}

\end{document}